\documentclass[english]{lni}

\usepackage{fancyhdr}
\usepackage[figurename=Fig., tablename=Tab., small]{caption}[2008/04/01]

\usepackage[ngerman]{babel}
\usepackage[utf8]{inputenc}
\usepackage{amsthm}
\usepackage{amssymb}
\usepackage{bibgerm}
\usepackage{graphicx}
\usepackage{color}
\usepackage{todonotes}
\usepackage{booktabs}
\usepackage[figuresright]{rotating}
\usepackage{url}

\newcommand{\myskip}[1]{}

\newcommand{\aID} {derived identity}
\newcommand{\aIDs}{derived identities}
\newcommand{\tFA}{two-factor authentication}
\newcommand{\X}{21}

\author{Daniel Träder$^2$ \, Alexander Zeier$^2$ \, Andreas Heinemann\footnote{
University of Applied Science Darmstadt, Fachbereich Informatik, Haardtring 100, 64295 Darmstadt \{daniel.traeder$\vert$alexander.zeier$\vert$andreas.heinemann\}@h-da.de. 
This work was funded by the Hessian Ministry of the Interior and Sports (HMdIS) within the “Round Table Cybersecurity@Hessen”.
}}

\title{Design and Implementation Aspects of Mobile Derived Identities\footnote{Extended Version of Daniel Träder, Alexander Zeier, and Andreas Heinemann, “Design and Implementation Aspects of Mobile Derived Identities”, in Open Identity Summit 2017}}

\begin{document}
\maketitle
\setcounter{footnote}{2} 

\begin{abstract}
With the ongoing digitalisation of our everyday tasks, more and more eGovernment services make it possible for 
citizens to take care of their administrative obligations online. This type of services 
requires a certain assurance level for user authentication. To meet these requirements, a digital identity 
issued to the citizen is essential. Nowadays, due to the widespread use of smartphones, mobile user authentication
is often favoured. This naturally supports two-factor authentication schemes (2FA). We use the term \emph{mobile derived
identity} to stress two aspects: a) the identity is enabled for mobile usage and b) the identity is somehow derived
from a physical or digital proof of identity.
This work reviews \X{} systems that support mobile derived identities. One subset of the considered systems is 
already in place (public or private sector in Europe), another subset is subject to research. Our goal is to identify prevalent design and implementation 
aspects for these systems in order to gain a better understanding on best practises and common views on mobile derived identities.
We found, that 
research prefers storing identity data on the mobile device itself whereas real world systems usually rely on
cloud storage. 2FA is common in both worlds, however biometrics  as second factor is the exception.
\end{abstract}


\section{Introduction}\label{intro}

Many online services require the unambiguous identification and authentication of users by the respective service providers. 
To reduce the risk of identity theft, two-factor authentication (2FA) schemes are more and more in place. The omnipresence of smartphones \cite{statistisches_bundesamt_smartphone_2016}
make these devices a natural choice for the second factor (possession), e.g.\ by using the embedded SmartCard or a personalised app, cf.\ \cite{terbu_one_2016}.

Online services provided by governments require special attention. For example, the European eIDAS regulation
\cite{eu_verordnung_2014} and the ISO-15408 \cite{iso15408}
define requirements for the technical realisation of eGovernment online authentication processes to fulfil predefined assurance levels. 
On the provider side, services are assigned to these assurance levels. This way, it can be ensured, that certain security and data sensitive processes can only be proceeded online if the authentication of a user and the corresponding online service operate on the same assurance level.

One important aspect is how the enrolment of a user's digital identity takes place, meaning the initial secure association of the identity data with a real person. If a proof of identity (digital or physical),  that was already verified by a trusted
third party, is at hand, we call this new identity \emph{digital derived identity}.
Provided that this digital derived identity can be used for mobile authentication, we call this
a \emph{mobile derived digital identity}.

\begin{figure}[htb]
  \centering
   \includegraphics[scale=0.7]{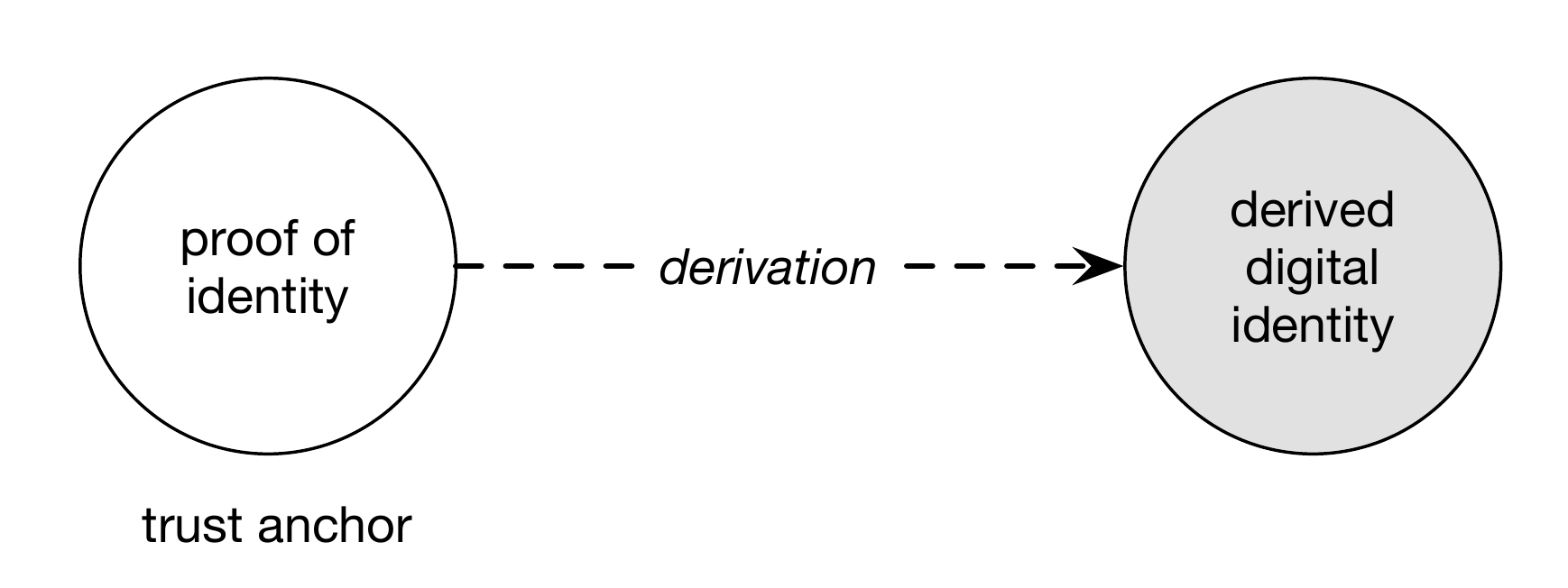}
  \caption{Derived digital identity}
  \label{fig:IDabID}
\end{figure}

Derived digital identities are characterised by the fact that they are linked to the 
proof of identity from which they were derived. This proof of identity is acting as a trust anchor, which is illustrated in Figure~\ref{fig:IDabID}.

The concept of digital derived identities leads to a number of interesting questions in terms of design and implementation:

\begin{itemize}
    \item How does the derivation process from the initial identity takes place?
    \item Where is the \aID{}\footnote{Further used as a short form for \emph{digital} derived identity.} stored and how is the access secured?
    \item How often can a \aID{} be generated?
\end{itemize}

During the course of this work, \X{} publications from research on mobile derived identities as well as systems already in place were analysed and essential design and implementation aspects were gathered. This paper provides an overview of our findings.

The remainder of this paper is structured as follows: First, we give an overview of related work (Section \ref{related}). Next, Section \ref{design} covers important aspects that we identified regarding the design and implementation of mobile derived identities. Based on these aspects more details on implementations are presented in Section \ref{solutions} before we summarise and conclude this work (Section \ref{conclusion}).

\section{Related Work}
\label{related}

Kubach et al. \cite{kubach_ssedic_2015} investigate mobile identity solutions in Europe, both in the public and private sector. Their work provides a deeper technical insight into some of the solutions for mobile identities. Gemalto \cite{gemalto_national_2014} provides an overview of public sector mobile eID solutions.
The \emph{Dutch Institute for Public Administration} \cite{pblq_international_2015} looks at European eID solutions with a focus on financial aspects and is limited to few mobile solutions.

To our knowledge, this is the first work that systematically investigates design and implementation aspects of \emph{mobile derived} digital identities within European countries as well as relevant research contributions in this field.



\section{Design and Implementation Aspects}
\label{design}

The realisation of mobile derived identities requires a proof of identity (a previous issued digital identity or a physical ID card/document) as well as a mobile device. The smartphone is a canonical representative of a mobile device. However, other \emph {wearables} \cite{mann-def-wearable} 
such as smart watches, fitness trackers etc.\ are also conceivable, as long as they have the possibility to connect directly or indirectly to an online service.

In the remainder of this section we elaborate more on the important design and implementation aspects we identified.

\textbf{Storage:} The foremost aspect addresses the storage location of the mobile derived identity.
There are two options: local on the mobile device itself or remote in the \emph{cloud}.
If the \aID{} is stored locally, special precautions must be taken to prevent unauthorised access.
Secure storage (e.g.\ a Secure Element \cite{se2017}) 
or SmartCards are typical options here).
Storing the derived identity in the cloud requires a cryptographic protection of the identity.
Access must be granted only by means of the owner of the derived identity. Typically,
a private key of an asymmetric cryptographic scheme is used in this context.

\textbf{Registration and Enrolment:} For the initial generation of a derived identity and its registration with an identity management system,
a process in the real world and an online process
can be distinguihed. In both cases,
the user has to provide his own identity proof and his own mobile device.
A real world registration process would be a visit to an authority which once again checks the authenticity of the identity proof and then generates the \aID{} and places it on the mobile device. 

For an online process further hardware (e.g.\ a card reader, a NFC reader, a camera) is often needed to verify the authenticity of the identity proof.

\textbf{Authentication:} The availability of a mobile device usually leads to the usage of a two-factor authentication scheme 
when accessing an online service. In addition to the possession and exclusive access to the device 
(as a first factor), the second factor varies depending on the capabilities of the device. Two possibilities
are the knowledge of a secret (e.g. PIN) or the use of biometric features (e.g. fingerprint, voice, iris etc.) which
require biometric sensors on the device.

\textbf{Derivation:} A last aspect of the initial generation of a \aID{} is the number of possible
derived identities in existence, i.e. is it possible to use the initial identity proof multiple times
to derive an identity (e.g. to place it on different mobile devices)? 
If several \aIDs{} are supported, one needs to specify whether these derived identities are distinguishable from each other and whether they have to be revoked individually or as a whole in case of loss or theft. Similarly, the technical requirements
of the mobile devices have to be specified accurately to secure access to the derived identities on all devices in the same manner. For example, a \aID{} must be stored equally secure on a smart watch as on a smartphone with a SIM card.

\section{Current Implementations of Mobile Derived Identities}
\label{solutions}

\begin{sidewaystable}
  \centering
  \includegraphics[scale=0.79]{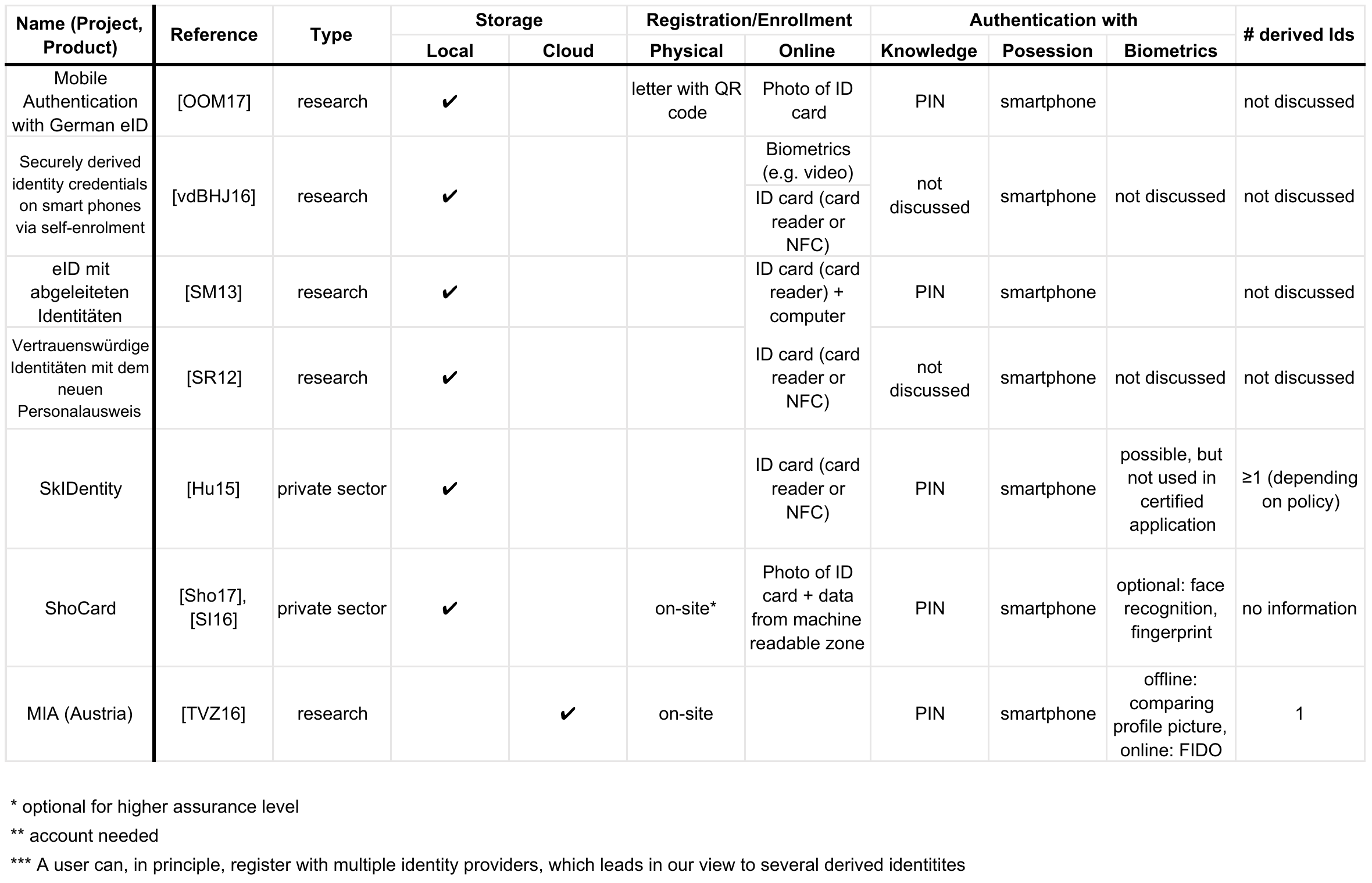}
  
  \caption{Realisation of mobile derived identities}
  \label{tab:overview}
\end{sidewaystable}

\begin{sidewaystable}
  \centering
  \includegraphics[scale=0.79]{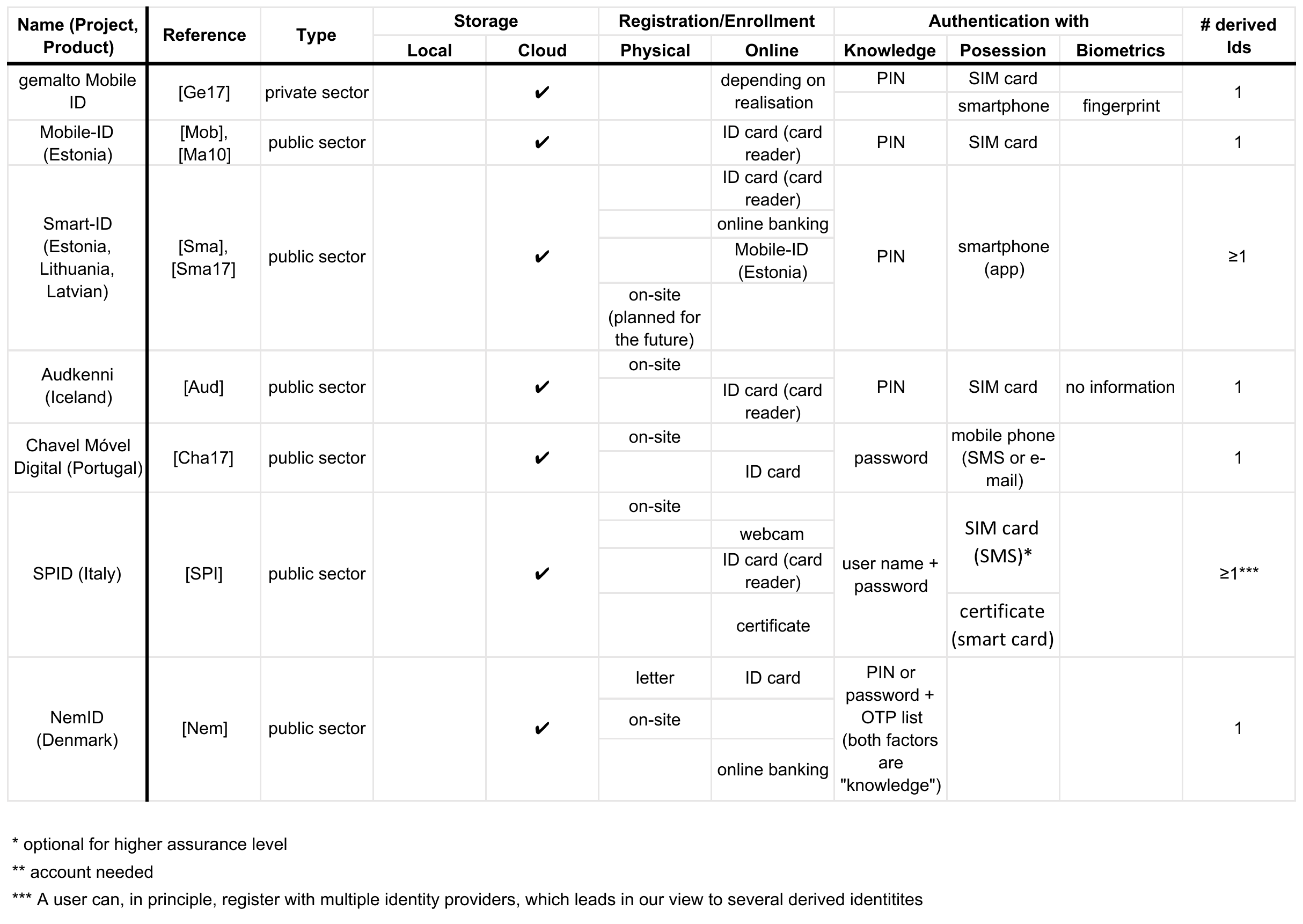} 
  
  \caption{Realisation of mobile derived identities (continued)}
  \label{tab:overview2}
\end{sidewaystable}

\begin{sidewaystable}
  \centering
  \includegraphics[scale=0.79]{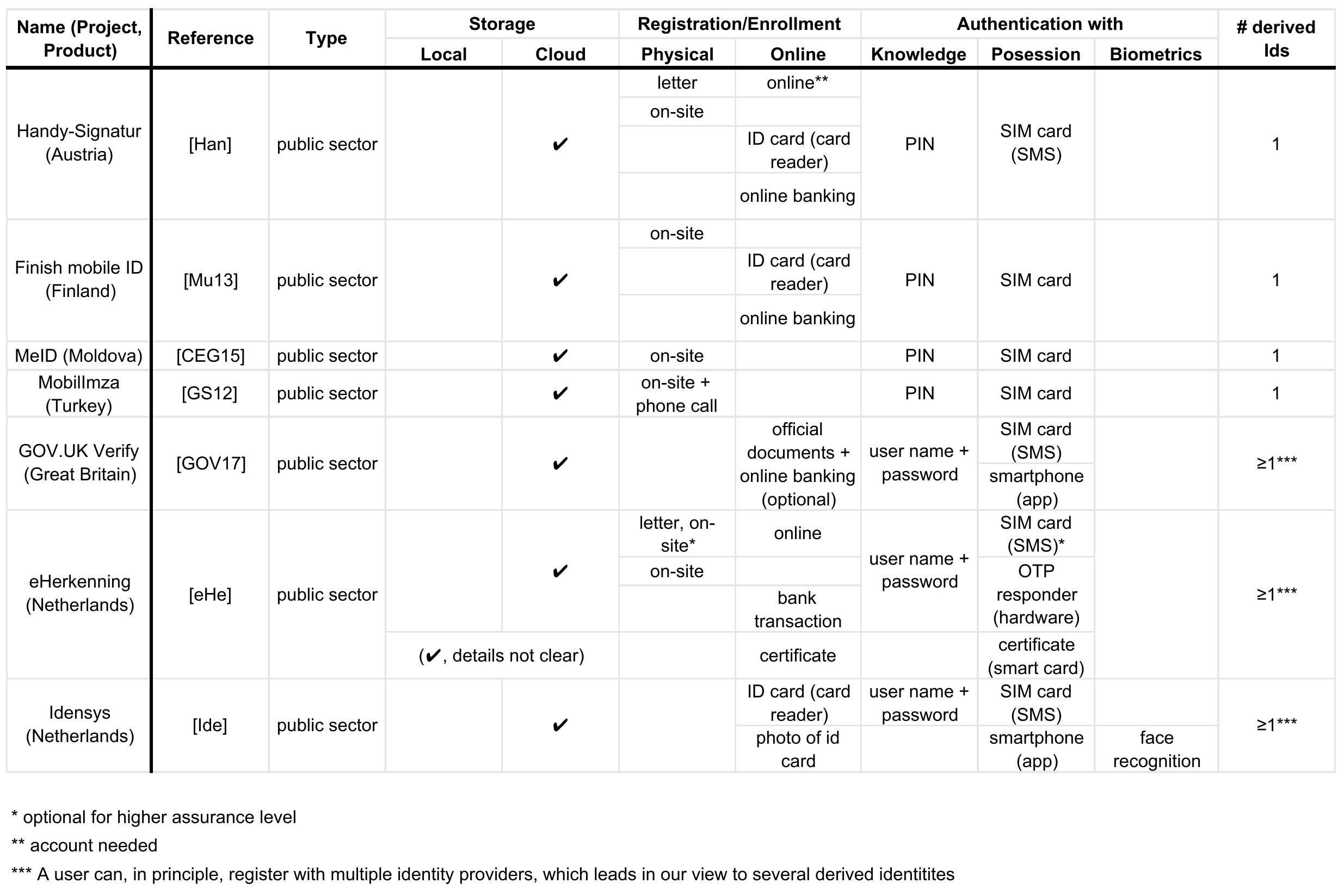} 
  
  \caption{Realisation of mobile derived identities (continued)}
  \label{tab:overview3}
\end{sidewaystable}

The proposals and systems investigated in this paper are summarised in table \ref{tab:overview} based on the aspects 
discussed in section \ref{design}.

We notice, that many of the systems in use in European countries store the \aID{} in the cloud, while scientific papers prefer local storage. 

Both, offline/real world and online registration are supported in most cases. The \tFA{} is present almost everywhere through knowledge and possession. Biometric readouts are rarely supported  and even then only optional. Likewise, the generation of several derived identities is the exception.

We provide further details on individual systems below. The \textit{Registration and Enrolment} and \textit{Storage Location and Usage of Derived Identities} aspects are considered, since their implementations show remarkable differences. 

\subsection{Storage Location and Usage of Derived Identities}
\label{storage}

The majority of cloud storage solutions (gemalto Mobile ID \cite{gemalto}, Mobile ID (Estonia) \cite{mobiil-id}, Audkenni (Iceland) \cite{audkenni}, Finish mobile ID \cite{murphy_finnish_2013}, MeID (Moldova) \cite{center_of_electronic_government_moldova_nodate}, MobilImza (Turkey) \cite{gsma_mobile_identity_team_mobile_2012}) use the SIM card of a user's phone to store a private key as a credential while the identity itself is stored on a server. In this case, if a user wants to authenticate against an online service, a signing request is sent to the user via the mobile operator. The user gets prompted with a message asking him if he wants to accept the execution of the initiated action. If he does so, the message gets signed in the background and sent back to 
the service provider, which then checks the validity of the signature. If successful, the user is authenticated.

Other cloud storage solutions (eHerkenning (Netherlands) \cite{eherkenning}, Chavel Móvel Digital (Portugal) \cite{chave}, SPID (Italy) \cite{spid}, NemID (Denmark) \cite{nemid}, Handy-Signatur (Austria) \cite{handy-signatur}, gemalto Mobile ID, Idensys (Netherlands) \cite{idensys}, GOV.UK Verify (Great Britian) \cite{gov.uk}) store no additional data on the mobile device and make use of the user's phone, or more specifically their SIM card, as a second factor by sending a SMS with a TAN code to the user which he then types into an input field in the browser, where the authentication process was initiated. With some solutions, the TAN can also be received by a smartphone application(Smart-ID (Estonia, Lithuania, Latvian) \cite{smart-id}, GOV.UK Verify) or e-mail\footnote{Here, any other device able to access the user's e-mail is sufficient.} (Chavel Móvel Digital (Portugal)). In case of Idensys, gemalto Mobile ID and MIA (Austria) biometric features (face, fingerprint) can be used if a user has installed a specific application on his device.

Locally stored derived identities are encrypted and saved on the device and the corresponding key is stored in its secure element.
In order to access the identity (e.g. during authentication) the user has to provide a PIN code in most cases. In addition, 
ShoCard gives the option to use face recognition or fingerprint analysis to access the identity. ShoCard is also the 
only solution that makes use of a blockchain to store a signed hash of the user's identity. This way a 3rd party can check the validity of the identity by comparing it to the stored hash in the blockchain.
However, the identity itself is stored locally.

\subsection{Registration and Enrolment}
\label{enrolment}

Most systems in use offer offline/real world registration. This on-site enrolment offers the possibility to enrol locally either at an authority office, a bank branch, or at an office of a mobile operator. This kind of enrolment is possible with ShoCard \cite{shocard}, MIA (Austria) \cite{terbu_one_2016}, Audkenni (Iceland), Chavel Móvel Digital (Portugal), SPID (Italy), NemID (Denmark), Handy-Signatur (Austria), Finish mobile ID (Finland), MeID (Moldova), MobilImza (Turkey) and eHerkenning (Netherlands) \cite{eherkenning}. Smart-ID (Estonia, Lithuania, Latvian) \cite{smart-id} plans to introduce on-site enrolment. 

The online enrolment is mostly performed with the help of an ID card with online capabilities. For the communication between the infrastructure provided by the government and the ID card, a card reader is indispensable. This is provided by many solutions: Idensys (Netherlands), Handy-Signatur (Austria), Finish mobile ID (Finland), SPID (Italy), Mobile-ID (Estonia) \cite{mobiil-id}, Smart-ID (Estonia, Lithuania, Latvian), Audkenni (Iceland), SkIDentity \cite{huhnlein_skidentity_2015}, \cite{van_den_broek_securely_2016}, \cite{schroder_eid_2013} and \cite{schmidt_vertrauenswurdige_2012}. 

Other solutions provide an enrolment without the need of a card reader but only the ID card. Chavel Móvel Digital (Portugal) and ShoCard make use of the machine readable zone (MRZ) printed on every card. The MRZ is read out with a camera. Next, the data is used as input for the generation of the
derived identity. 

GOV.UK Verify (Great Britain), eHerkenning (Netherlands), Finish mobile ID (Finland), Handy-Signatur (Austria), NemID (Denmark) and Smart-ID (Estonia, Lithuania, Latvian) cooperate with financial institutions for the registration based on previous online-banking enrolment. In general there are two methods to register based
on online-banking: A small amount of money is transferred to the bank account of the registration provider. The proof of identity is given to the person who commissioned the transaction. To enrol with the second method, the customer of a bank has to answer questions about past bank transactions. The questions are about the transferred and received money on the customer account, e.g. the amount and time of a specific transaction. In this solution, the knowledge about past transactions proofs the identity of the account owner. SPID (Italy) offers a video identification. During a video chat an employee of the registration provider performs the identification. Video identification is suggested by \cite{van_den_broek_securely_2016} as well.

In addition, a combination of offline and online actions that need to be carried out by a user already exist. \cite{otterbein_mobile_2017} and ShoCard make use of a photo taken from the physical ID card or just from the machine readable zone on the card. 
Whereas eHerkenning (Netherlands), NemID (Denmark) and Handy-Signatur (Austria) make uses of the serial number which has to be entered by the user in an online form. In all cases, a letter will be sent to the registered address of the card holder. This letter holds private information to complete the enrolment process.

\section{Conclusion and Outlook}
\label{conclusion}

We have investigated prevalent design and implementation aspects of mobile derived identities. We studied
\X{} research contributions and systems used in practice. The most prominent design criterion is
the storage location (local or cloud) of a mobile derived identity.
It can be seen that most of the public sector solutions in European countries make use of a cloud storage, 
while all (except \cite{terbu_one_2016}) scientific works considered prefer a local storage on a personal device for the identity.

Most of the public sector solutions offer an on-site registration and enrolment. Two-factor authentication is implemented in all considered systems. For NemID both factors are know\-ledge. The use of biometric features 
as a second factor is the exception.

For the sake of completeness, we note, that in the context of eGovernment online services legal requirements
must be taken into account, see \cite{eu_verordnung_2014} for example, affecting the design decisions.

As a next step, we will take a closer look at the different assurance levels and review whether a \aID{} shows the same assurance level as the identity from which it was derived (this applies only for digital identities as an identity proof). In addition, we want to look closer into usability issues of the various realisations, and how a
citizen must deal with loss or damage of their mobile device.



\nocite{otterbein_mobile_2017}
\nocite{neven_d32.2_2012}
\nocite{rosnagel_futureid_2012}
\nocite{huhnlein_skidentity_2015}
\nocite{shocard}
\nocite{sita_travel_2016}
\nocite{van_den_broek_securely_2016}
\nocite{schroder_eid_2013}
\nocite{eichholz_verfahren_2009}
\nocite{eherkenning}
\nocite{martens_electronic_2010}
\nocite{audkenni}
\nocite{chave}
\nocite{spid}
\nocite{nemid}
\nocite{handy-signatur}
\nocite{murphy_finnish_2013}
\nocite{center_of_electronic_government_moldova_nodate}
\nocite{terbu_one_2016}
\nocite{gsma_mobile_identity_team_mobile_2012}
\nocite{gov.uk}
\nocite{idensys}
\nocite{gemalto}
\nocite{mobiil-id}
\nocite{schmidt_vertrauenswurdige_2012}
\nocite{smart-id}
\nocite{smart-id-technical}

\bibliographystyle{lni}
\bibliography{literature} 

\end{document}